\documentstyle[11pt,paspconf,epsf]{article}

\begin{document}

\title{GHRS Monitoring of the Outflowing Material\\
       in NGC~4151}

\author{Ray J. Weymann}
\affil{The Observatories of the Carnegie Institution of Washington,\\ 
       813 Santa Barbara St., Pasadena, CA 91101, \\ e-mail: rjw@ociw.edu}

\author{Simon L. Morris, Meghan E. Gray, John B. Hutchings}
\affil{Dominion Astrophysical Observatory, National Research Council,\\
       5071 West Saanich Road, Victoria, B.C., V8X 4M6, Canada,\\ 
       e-mail: simon@dao.nrc.ca, gray@dao.nrc.ca, hutchings@dao.nrc.ca}

\begin{abstract}
We present the results of a GHRS program to monitor the absorption
lines in the spectrum of the Seyfert 1 galaxy NGC~4151 caused by
outflowing gas from the nucleus.  Although we see subtle changes over
the four year period in the GHRS spectra of the broader of the
absorption features, the wavelength constancy of all the features is
remarkable. The limits on the secular acceleration suggest that either
(1) The absorbing clouds are well beyond the broad emission line
region, or (2) The clouds are experiencing significant drag from an
intercloud medium. The exception to this constancy occurred during one
of the epochs of our monitoring when a broad shallow C~IV trough
appeared at an outflow velocity of 3750~km~s$^{-1}$ and then
subsequently disappeared.
\end{abstract}


\section{Observations} \label{sec-obs}

\begin{figure}[ht]
\plotfiddle{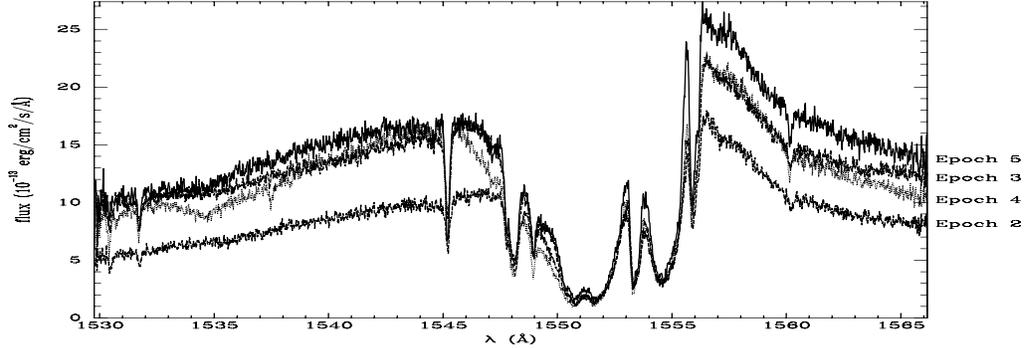}{4cm}{90}{60}{30}{225}{-36}
\caption{C~IV absorption in the GHRS spectra of NGC~4151  \label{fig-weymannr1.eps}}
\end{figure}

NGC~4151 is one of the nearest, brightest and most studied
of the classical Seyfert galaxies and is considered the
prototype Type 1 Seyfert. In addition to its broad permitted lines, it
has been recognized for many years  that there were also
blueward-displaced absorption features present, indicating the presence
of outflowing gas.  GHRS observations of the nucleus of NGC~4151 were
made over five epochs spanning 1992 June 22--27 (Epoch 1), 1992 July 4
(Epoch 2), 1994 January 3 (Epoch 3), 1994 October 28--29 (Epoch 4), and
1996 March 11 (Epoch 5). The flux calibrated observations of the C~IV 
absorption are shown in figure~\ref{fig-weymannr1.eps}

\section{Profile Fitting and Limits on Kinematic Variability} \label{sec-prof}

\begin{figure}[ht]
\plotfiddle{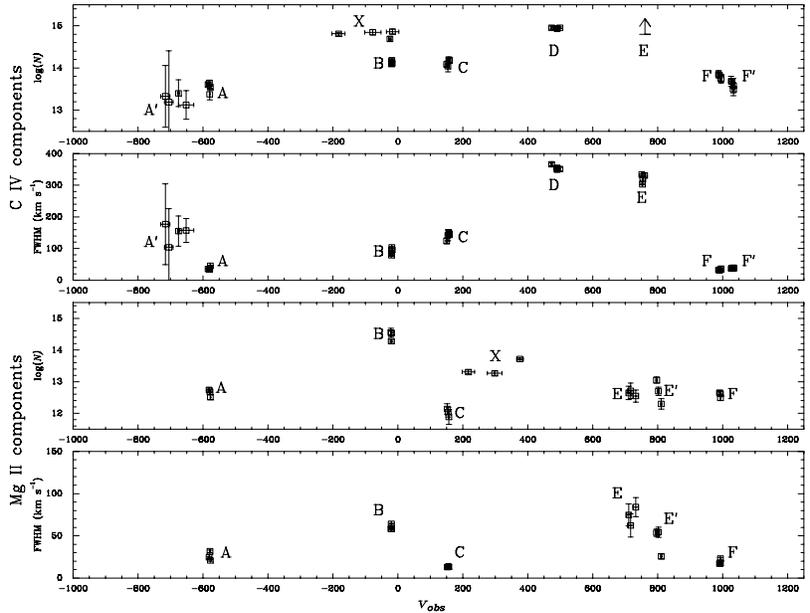}{8cm}{90}{50}{50}{225}{-36}
\caption{Column density and FWHM vs velocity for C~IV and Mg~II.  The FWHM 
values for Component X do not appear on  this scale. \label{fig-weymannr3.eps}}
\end{figure}

Normalized spectra were produced by dividing the flux data and their
corresponding error vectors by a continuum + emission line fit.  The
\begin{figure}[ht]
\plotfiddle{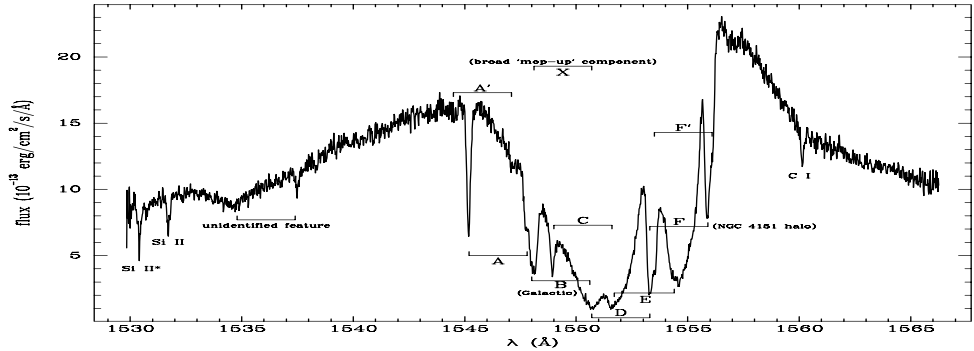}{4cm}{90}{60}{30}{225}{-36}
\caption{C~IV component IDs  \label{fig-weymannr2.eps}}
\end{figure}
absorption line strengths (refer to Figure~\ref{fig-weymannr2.eps} for line
IDs) were measured using the `specfit' task in IRAF (\cite{kriss94}),
using combinations of {\em tauabs} models (Gaussian function in optical
depth) and {\em labsorp} (Gaussian in equivalent width).
Figure~\ref{fig-weymannr3.eps} contains plots of $\log N$ vs $v_{obs}$ and
$\log N$ vs FWHM for both the C~IV components and the Mg~II
components.

In Epoch 4, a wide trough was observed at $\sim$1534.5{\AA}.  It is
almost certainly not the result of a defect on the photocathode. It
remains a possibility that the apparent absorption is in fact a
combination of an offset emission feature (related to the satellite
features of \cite{ulrich85}), and a poor fit to the resulting line and
continuum.  However, the wavelength of the \cite{ulrich85} L1 feature
is 1523.5{\AA}, significantly bluer than is needed to easily explain
this absorption.

\section{Limits on secular acceleration of the absorbing clouds} \label{sec-accel}

Inspection of Figures~\ref{fig-weymannr1.eps} and
\ref{fig-weymannr3.eps} shows that over the time covered by the
observations at Epochs 2--5, while there have been some small changes
in the absorption profiles (notably the appearance of the high velocity
transient at epoch 4), the gross properties have been remarkably
stable. {\it In particular, we find no evidence for any secular change
in velocity in the well defined A and C components.} A fairly
conservative limit on any secular steady acceleration for these is
about 1$\times$10$^{-3}$~cm~sec$^{-2}$.

We consider two naive interpretations of this limit on the
acceleration:

\noindent
(1) The clouds are undergoing free radiative acceleration.  For an
order of magnitude estimate, we can consider them to be optically thin
to both lines and continua. With some reasonable assumptions about
abundance and geometry, the limit on the secular free acceleration
would require the cloud to be at a distance of about
1.7$\times$10$^{18}\times$D$_{10}$ cm, where D$_{10}$ is the distance
from us to NGC~4151 in units of 10 Mpc.  This is far beyond the broad
emission line region, which, on the basis of reverberation studies
(\cite{clav90}) is of order 10$^{16}$ cm.

\noindent
(2) In the second naive model, we assume that a cloud is exposed to
radiative acceleration, as estimated above, with a distance from the
central source comparable to the size of the broad emission line region
(i.e., about 10 cm~s$^{-2}$), but that drag against a hot confining
medium counteracts the radiative acceleration. A static medium with a
density as low as about 10$^2$ cm$^{-3}$ would exert enough drag on the
clouds to prevent detectable acceleration.  However, a static medium seems very
unlikely, since its drag would prevent the clouds from reaching
their observed terminal velocities.  A more likely scenario would
involve an expanding and accelerating wind at higher density with a
small velocity differential between the cloud and the accelerating wind
(c.f. the discussion in \cite{beg91}).  There are two limiting
situations: (a), where we assume that the acceleration commences at
the scale associated with the C~IV broad emission line region (BELR).
Then the cloud must be well beyond the BELR, as we inferred for the
ballistic radiatively accelerated scenario. Or (b), where we require
that the absorbing cloud be at, or only very slightly beyond, the C~IV
BELR, as is generally assumed. Then the acceleration must commence at a
distance very much smaller than the BELR in order that we currently not
be able to detect the acceleration. 

Implicit in both these scenarios is the assumption that we are
observing the same material over the duration of our observations,
rather than a flow pattern whose dynamical properties have not changed
over several years. 

\section{Summary} \label{sec-sum}

The observations discussed here have some interesting similarites with
 the much more luminous Quasars. While outflow from Seyfert nuclei does
not appear to be uncommon, the outflow
velocities are generally restricted to velocities of order 3500
km~s$^{-1}$ or less -- i.e., about 1/10 the velocities in the luminous
BALQSOs. If the dominant
accelerating mechanism is radiative and optically thin material is
accelerated from rest beginning at some radius $r_o$ upon being exposed
to luminosity $L$, then the terminal velocity should scale as $
V_{\infty} \sim (L/r_o)^{1/2}$. It appears that the range of ionization
parameters characterizing the broad emission line region of AGNs is
rather similar over the very large range of optical luminosities
spanned by AGN, so that $L/(n_e \times r_o^2)$ is roughly constant.
Moreover, most models infer rather similar electron densities over this
range, so that one expects $ r_o \sim L^{1/2}$ which leads to the
scaling law $V_{\infty} \sim L^{1/4}$. The difference in luminosities
between the rather modest Seyfert luminosity of NGC~4151 and the most
luminous BALQSOs is about 10 magnitudes, implying a range in terminal
velocities of about 10, which is roughly what is observed. Almost no
work has been done to explore the terminal velocities of BAL-type
objects near what is generally considered to be the dividing line
between QSOs and Seyferts (ie. about M = -23), if there are such
objects, and it would be of interest to see if they fall along such a
sequence.

\acknowledgments

We are grateful to Gerry Kriss for help with using his software package
`specplot'. RJW acknowledges support through NASA contract
NAS 5-30101 and NSF grant AST 90-05117.

\end{document}